\documentclass{emulateapj}
\usepackage{apjfonts}
\usepackage{amsmath}
\newcommand{\sj}{{\scriptscriptstyle (j)}}
\newcommand{\tot}{\rm{tot}}
\newcommand{\sub}{\rm{sub}}
\newcommand{\obs}{\rm{obs}}
\newcommand{\dep}{\rm{dep}}
\newcommand{\dur}{\rm{dur}}
\shorttitle{$E_p$--$E_{iso}$ Correlation in a Multiple Subjet Model}
\shortauthors{Toma, Yamazaki, \& Nakamura}
\begin{document}
\title{$E_p$--$E_{iso}$ Correlation
in a Multiple Subjet Model of Gamma-Ray Bursts
}
\author{Kenji Toma$^{1}$, Ryo Yamazaki$^{2}$
and Takashi Nakamura$^{1}$}
\affil{
$^{1}$Department of Physics, Kyoto University,
Kyoto 606-8502, Japan
\\
$^{2}$Department of Earth and Space Science,
Osaka University, Toyonaka 560-0043, Japan
}

\email{
toma@tap.scphys.kyoto-u.ac.jp
}

\begin{abstract}

We perform Monte Carlo simulations
to study $E_p$--$E_{iso}$ correlation in the context of a multiple 
subjet model (or inhomogeneous jet model) for gamma-ray bursts (GRBs),
X-ray--rich GRBs (XRRs), and X-ray flashes (XRFs).
For a single subjet, we find that $E_p \propto {E_{iso}}^{0.4}$
for large viewing angles.
For the multiple subjet model in which all the subjets have the same intrinsic
properties, off-axis events 
show $E_p \propto {E_{iso}}^a$ with $0.4 < a < 0.5$.
If the intrinsic properties of the subjets are distributed so that
on-axis emission of each subjet follows a correlation 
$E_p \propto {L_{iso}}^{1/2}$,
we obtain the Amati correlation ($E_p \propto {E_{iso}}^{1/2}$) 
over three orders of magnitude in $E_p$.
Although the scatter around the Amati correlation is large in the simulation,
the results are consistent with the observed properties of 
GRBs with known redshifts
and the BASTE GRBs with pseudo redshifts derived from
the lag-luminosity correlation.
We also calculate the event rates, the redshift distributions, and the $T_{90}$ duration
distributions of GRBs, XRRs, and XRFs which can be detected by {\it HETE-2},
assuming that the source redshift distribution is in proportion to the cosmic star
formation rate.
It is found that the event rates of three classes are comparable, 
that the average redshift of the XRRs is a little larger than those of the GRBs and the XRFs,
and that short XRRs arise when a single subjet is viewed off-axis or viewed on-axis with 
slightly high redshift.

\end{abstract}


\section{Introduction}
\label{sec:intro}

{\it HETE-2} observations have provided strong evidence
that softer and dimmer gamma-ray bursts (GRBs) smoothly
extend to X-ray flashes (XRFs) through an intermediate class
of events called
X-ray--rich GRBs (XRRs).
For events with known redshifts and well observed spectra,
the rest-frame spectral peak energy $E_p$ and
the ``bolometric'' isotropic-equivalent $\gamma$-ray energy $E_{iso}$
have a strong correlation, i.e., $E_p \propto {E_{iso}}^{1/2}$
\citep{amati02}.
This $E_p$--$E_{iso}$ correlation, called 
the Amati correlation, has recently been
extended down to lower energies
characteristic of XRFs \citep{lamb04}.
Since various observed quantities other than
the Amati correlation also distribute continuously among
GRBs, XRRs, and XRFs \citep{saka05}, it is
strongly suggested that these three classes
are related phenomena.

While many different models have been proposed for XRFs
\citep[see][and references therein]{granot05},
we have proposed the ``off-axis model'' \citep{yama02,yama03}
in which XRFs are the usual GRB jets viewed from an off-axis
viewing angle \citep[see also][]{woods99}. 
When the jet is observed off-axis, the emitted photons
are out of the beaming
cone and less blueshifted than photons emitted along the jet axis,
so that the events look like XRFs.
It has been shown that
the viewing angle of the jet is the key parameter
to understand the various properties of the GRBs and
that the luminosity-variability/lag/width correlations
might be naturally derived in the framework of 
off-axis models \citep{ioka01}.

As for the Amati correlation,
\citet{yama04a} computed $E_p$ and $E_{iso}$
using the uniform jet model and found that the
results are compatible with the observations. 
They also found that $E_p \propto
{E_{iso}}^{1/3}$ in the smaller $E_{iso}$ regime.
\citet{eichler04} investigated the correlation
in an annular jet model,
and derived that if the viewing angles are within the
annulus, $E_p \propto {E_{iso}}^a$
with $1/3 < a <1/2$, which is compatible with
the observations.
Compared with our uniform jet model, in the annular jet model
the energy is large due to the emissions from
widely distributed segments with similar viewing angles.
\citet{eichler04} also
anticipated that multiple discrete emissions could have
the same effect.

The off-axis jet model has recently
been improved to include short GRBs \citep{yama04b}
as a unified model, where
the GRB jet is not uniform but made up of multiple subjets or
multiple emission patches.
This is an extreme case of an inhomogeneous jet model
\citep{nakamura00,kumar00}.
The crucial parameter is the multiplicity, $n_s$, of
the subjets along the line of sight. If $n_s \geq 2$,
the burst looks like a long GRB, and if $n_s = 1$
the burst looks like a short GRB, while if $n_s = 0$
the burst is an off-axis event for
all the subjets and looks like an XRF or an XRR.
We also found that the unified model may explain the bimodal
distribution of the $T_{90}$ durations of BATSE GRBs
\citep{toma05}.

In this paper, we examine $E_p$--$E_{iso}$ correlation in the
multiple subjet
model to show that the unified model is consistent with
the observations of $E_p$ and $E_{iso}$.
This paper is organized as follows.
In \S~\ref{sec:model}, we describe our multiple subjet model for prompt
emissions.
First, the $E_p$--$E_{iso}$ correlation for a single subjet is discussed
in \S~\ref{sec:single}, and then we discuss the results of Monte Carlo
simulations
in the multiple subjet model in \S~\ref{sec:multi}.
Section~\ref{sec:dis} is devoted to discussion.

\section{Prompt emission model}
\label{sec:model}

Let us suppose that $N_{\tot}$ subjets with opening half-angle
$\Delta\theta_{\sub}^{\sj}$ are launched from the central engine of
GRB randomly in time and directions and that the whole jet with opening
half-angle $\Delta\theta_{\tot}$ consists of these subjets.
We introduce the spherical coordinate system $(r, \vartheta, \varphi)$ in
the central engine frame, where the origin is the location of the
central engine, and $\vartheta=0$ is the axis of the whole jet.
The axis of the $j$th subjet $(j=1,\cdots, N_{\tot})$ is denoted by
$(\vartheta^{\sj}, \varphi^{\sj})$.
If the direction of the observer is
given by $(\vartheta_{\obs}, \varphi_{\obs})$,
the viewing angle of the $j$th subjet from the line of sight is
\begin{equation}
\theta_v^{\sj} = \cos^{-1} [\sin\vartheta_{\obs} \sin\vartheta^{\sj}
\cos(\varphi_{\obs}-\varphi^{\sj}) +
\cos\vartheta_{\obs}\cos\vartheta^{\sj}].
\label{eq:viewing}
\end{equation}

For emission model of each subjet, we use the same formulations and notations
as used in \citet{yama03}.
Let us use another spherical coordinate system $(r, \theta, \phi)$ in the 
central engine frame, where the origin is the location of the central engine, 
and $\theta =0$ is the line of sight.
We adopt an instantaneous emission, at $t=t_0^{\sj}$ and $r=r_0^{\sj}$, of an
infinitesimally thin shell moving with the Lorentz factor $\gamma_{\sj}$.
Then one can obtain the formula of
the observed flux from the $j$th subjet with viewing angle $\theta_v^{\sj}$
at frequency $\nu$ and time $T$,
\begin{equation}
F^\sj_{\nu}(T) = \frac{2(1+z)r_0^{\sj}c A^{\sj}}{d_L^2}
\frac{\Delta\phi^{\sj}(T)
f^{\sj}[(1+z)\nu \gamma_{\sj} (1-\beta^{\sj} \cos\theta(T))]}
{\gamma_{\sj}^2 (1-\beta^{\sj} \cos\theta(T))^2},
\label{eq:flux}
\end{equation}
where $z$ and $d_L$ are the redshift and the luminosity distance 
of the source, respectively;
$f^{\sj}(\nu')$ and $A^{\sj}$ represent the spectral shape and the amplitude
of the emission in the comoving frame, respectively.
Here $T = 0$ is chosen as the time of arrival at the observer
of a photon emitted at the origin at $t=0$.
The set of points that emit photons observed at a given time $T$
is an arc (or a circle).
The functions $\theta(T)$ and $\Delta\phi^{\sj}(T)$ represent
an angular radius and a central angle of the arc, respectively:
\begin{align}
\cos\theta(T) &= \frac{c}{r_0^{\sj}} \left(t_0^{\sj} - \frac{T}{1+z} \right), 
\label{eq:theta} \\
\Delta\phi^{\sj}(T) &= 
\begin{cases}
\pi, \\
(\text{for}~ \theta_v^{\sj}<\Delta\theta_{\sub}^{\sj}~ {\rm and}~
0<\theta(T)<\Delta\theta_{\sub}^{\sj}-\theta_v^{\sj}), 
\\
\cos^{-1} \left(\frac{\cos\Delta\theta_{\sub}^{\sj}
-\cos\theta_v^{\sj} \cos\theta(T)}{\sin\theta_v^{\sj} \sin\theta(T)}\right), \\
(\text{for otherwise}).
\end{cases}
\label{eq:phi}
\end{align}
Equation (\ref{eq:theta}) can be rewritten by
\begin{equation}
1-\beta^{\sj} \cos\theta(T) = \frac{1}{r_0^{\sj}/c\beta^{\sj}}
\left(\frac{T}{1+z} - t_{\dep}^{\sj} \right),
\label{eq:theta2}
\end{equation}
where $t_{\dep}^{\sj} = t_0^{\sj} - r_0^{\sj}/c\beta^{\sj}$ is the departure time 
from the central engine of the $j$th subjet. 

The observed spectrum of GRBs is well approximated by the Band spectrum
\citep{band93}.
In order to have a spectral shape similar to the Band spectrum,
we adopt the following form of the spectrum in the comoving frame,
\begin{equation}
f^{\sj}(\nu') = 
\begin{cases}
(\nu'/{\nu'_0}^{\sj})^{1+\alpha_B^{\sj}} \exp(-\nu'/{\nu'_0}^{\sj}) \\
({\rm for}~\nu'/{\nu'_0}^{\sj} \leq \alpha_B^{\sj} - \beta_B^{\sj}), 
\\
(\nu'/{\nu'_0}^{\sj})^{1+\beta_B^{\sj}}
(\alpha_B^{\sj} - \beta_B^{\sj})^{\alpha_B^{\sj} - \beta_B^{\sj}}
\exp(\beta_B^{\sj} - \alpha_B^{\sj}) \\
({\rm for}~\nu'/{\nu'_0}^{\sj} \geq \alpha_B^{\sj} - \beta_B^{\sj},
\end{cases}
\label{eq:spectrum}
\end{equation}
where ${\nu'_0}^{\sj}$, $\alpha_B^{\sj}$, and $\beta_B^{\sj}$ are the break
frequency and the low- and high- energy photon index, respectively.

As a summary, 
equations (\ref{eq:flux}), (\ref{eq:phi}), (\ref{eq:theta2}), and (\ref{eq:spectrum})
are the basic equations to calculate the observed flux from each subjet, 
which depends on the following parameters:
$\theta_v^{\sj}$ (which is determined by $\vartheta^{\sj}$, $\varphi^{\sj}$, 
$\vartheta^{\obs}$, and $\varphi^{\obs}$ through equation (\ref{eq:viewing})),
$\Delta\theta_{\rm sub}^{\sj}$, $\gamma^{\sj}$, $t_{\dep}^{\sj}$,
$r_0^{\sj}$, $\alpha_B^{\sj}$, $\beta_B^{\sj}$, ${\nu'_0}^{\sj}$,
$A^{\sj}$, and $z$.
The whole light curve from the GRB jet is produced by the superposition
of the emissions from the subjets.

\section{$E_p$--$E_{iso}$ correlation for a single subjet}
\label{sec:single}

\begin{figure}
\plotone{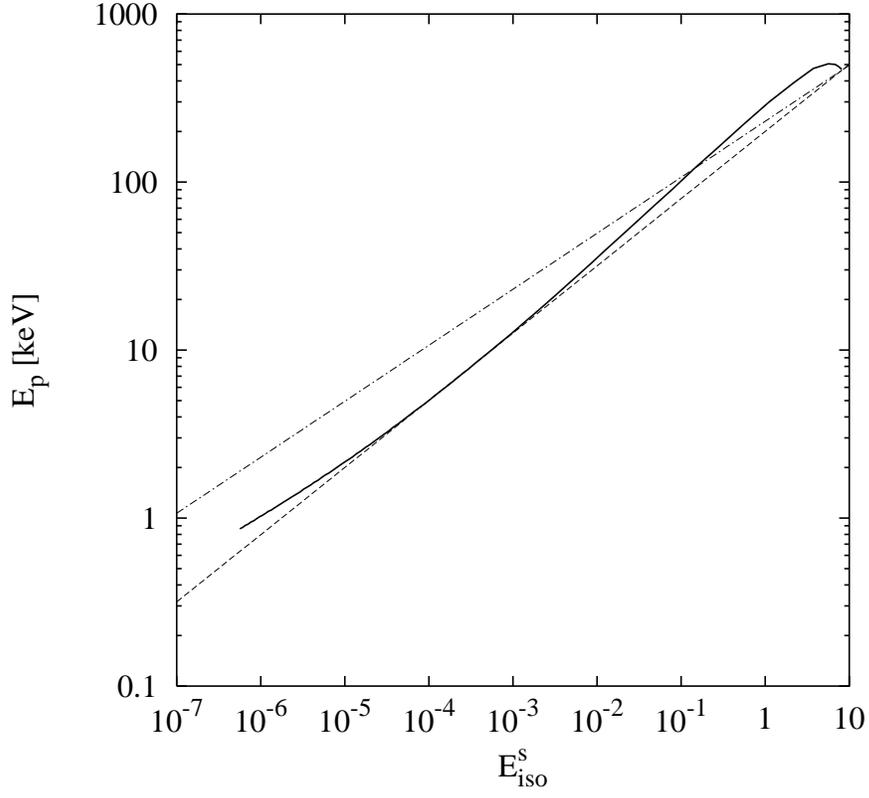}
\caption{
Correlation between the isotropic-equivalent energy $E_{iso}^s$
(in units of $2.8 \times 10^4 \pi A r_0^2$) and the spectral peak energy
$E_p$ for a single subjet.
${E_{iso}^s}{}^{0.4}$ ({\it dashed}) and ${E_{iso}^s}{}^{1/3}$ 
({\it dot-dashed}) lines are also shown.
}
\label{fig:single}
\end{figure}

Before examining $E_p$--$E_{iso}$ correlation for the multiple subjet
model,
it is instructive to calculate $E_p$--$E_{iso}$ correlation when a
single subjet is seen off-axis.
Using equations (\ref{eq:flux}), (\ref{eq:phi}), (\ref{eq:theta2}), 
and (\ref{eq:spectrum}) for $N_{\tot}=1$ and a given
$\theta_v$, we compute the peak energy of the time-integrated spectrum
measured in the cosmological rest frame, $E_p$, 
and the ``bolometric'' isotropic-equivalent energy,
$E_{iso}^s$, integrating over the $1-10^4$~keV range
in the cosmological rest frame.
Here the superscript $s$ of $E_{iso}^s$ means ``single''.
We adopt the following subjet parameters:
$\Delta\theta_{\sub} = 0.02$~rad, $\gamma=300$, $\alpha_B = -1$,
$\beta_B = -2.5$, and $\gamma h\nu'_0 = 350$~keV.
In Figure~\ref{fig:single}, we show $E_p$ and $E_{iso}^s$
(in units of $2.8 \times 10^3 \pi A {r_0}^2$)
for $0 < \theta_v < 0.1$~rad (the solid line).
The dashed and dot-dashed lines are ${E_{iso}^s}^{0.4}$ and 
${E_{iso}^s}^{1/3}$,
respectively.
We see that 
for $\theta_v > \Delta\theta_{\sub}$, as $\theta_v$ increases,
both $E_p$ and $E_{iso}^s$ decreases.
We focus on small $E_{iso}^s$ regime.
At first the $E_p$--$E_{iso}$ correlation approaches 
$E_p \propto {E_{iso}^s}^{0.4}$,
but for even larger $\theta_v$, $E_p \propto {E_{iso}^s}^{1/3}$.
This behavior is explained as follows.
Firstly the spectral peak energy scales as
$E_p \propto [1-\beta\cos(\theta_v - \Delta\theta_{\sub})]^{-1}$
because of the Doppler effect,
and for large $\theta_v$, $E_p \propto {\theta_v}^{-2}$.
Next we compute $E_{iso}^s$ by integrating equation (1) over $\nu$ and $T$,
and study its dependence on $\theta_v$. 
When $\theta_v$ is large but $E_p$ is in the $1-10^4$~keV range,
the integration over $\nu/(1+z)$
results in a constant depending on the Band spectral parameters and another
Doppler factor $ [1-\beta\cos\theta(T)]^{-1}$.
As for the integration with respect to $T$,
we change the variable from $T$ to $\theta(T)$, and obtain:
\begin{equation}
E_{iso}^s \propto
\int^{\theta_v+\Delta\theta_{\sub}}_{\theta_v-\Delta\theta_{\sub}}
\frac{\Delta\phi(\theta) \sin\theta d\theta}
{(1-\beta\cos\theta)^3}.
\label{eq:eiso}
\end{equation}
For large $\theta_v$, $\Delta\phi \simeq \Delta\theta_{\sub}/\theta_v$,
so that $E_{iso}^s \propto 
{\theta_v}^{-1}([1-\beta\cos(\theta_v-\Delta\theta_{\sub})]^{-2}
-[1-\beta\cos(\theta_v+\Delta\theta_{\sub})]^{-2}) \simeq
{\theta_v}^{-1} [1-\beta\cos(\theta_v-\Delta\theta_{\sub})]^{-2}
\propto {E_p}^{2.5}$.
When $\theta_v$ is even so large as $E_p \sim 1$~keV,
the integration over $\nu/(1+z)$ results in a factor 
$(1-\beta\cos\theta)^{1+\beta_B}$,
so that the same calculation gives us 
$E_{iso}^s \propto {\theta_v}^{-1} 
[1-\beta\cos(\theta_v-\Delta\theta_{\sub})]^{\beta_B} 
\propto {E_p}^3$ for $\beta_B = -2.5$.

For a single subjet, off-axis events obey $E_p \propto {E_{iso}}^{0.4}$
for small $E_p$ regime (but $E_p > 1$~keV).
The index of the $E_p$--$E_{iso}$ correlation, $a=0.4$, is obtained 
irrespective of the intrinsic subjet parameters 
$\Delta \theta_{\sub}$, $\gamma$, $t_{\dep}$, $r_0$, 
$\alpha_B$, $\beta_B$, $\nu'_0$, and $A$, 
as can be seen in the above derivation.

\section{$E_p$--$E_{iso}$ correlation in the multiple subjet model}
\label{sec:multi}

Let us perform Monte Carlo simulations to derive
$E_p$--$E_{iso}$ correlation in the multiple subjet model.
For simplicity, we generate one GRB jet with opening half-angle
$\Delta\theta_{\tot} = 0.3$~rad and random
5000 lines of sight of the observer
with $0<\vartheta_{\obs}<0.35$~rad according to the probability
distribution of $\sin\vartheta_{\obs}~d\vartheta_{\obs}~d\varphi_{\obs}$.
Then, for each observer, we calculate the peak energy
of the time-integrated spectrum measured in the cosmological rest frame,
$E_p$,
and the ``bolometric'' isotropic-equivalent energy, $E_{iso}$,
integrating over the $1-10^4$~keV range in the cosmological rest frame.
The departure time of each subjet $t_{\dep}^{\sj}$ is assumed to be
homogeneously random between $t=0$ and $t=t_{\dur}$,
where $t_{\dur}$ is
the active time of the central engine measured in its own frame,
and $t_{\dur} = 20$~s is adopted.
The central engine is assumed to produce
$N_{\tot}=350$ subjets following the angular
distribution function
\begin{equation}
\frac{dN}{d\Omega}
\propto
\begin{cases}
1, & 0<\vartheta^\sj<\vartheta_c, 
\\
(\vartheta^\sj/\vartheta_c)^{-2}, & \vartheta_c<\vartheta^\sj<\vartheta_b,\cr
\end{cases}
\label{eq:jetP}
\end{equation}
where $\vartheta_b = \Delta\theta_{\tot} - \Delta\theta_{\sub}$, and
$\vartheta_c = 0.03$~rad.
This corresponds to the universal structured jet model
\citep[see][]{rossi02,zhang02a}.
The angular distribution of the subjets in our simulations is shown
in Figure~\ref{fig:jetP}. The solid circle describes each subjet
and the dashed circle describes the whole jet. 
The meaning of plus sign will be discussed later.
We assume that all the subjets have the same values of the following 
parameters:
$\Delta\theta_{\sub}^{\sj} = 0.02$~rad, $\gamma_{\sj} = 300$,
$r_0^{\sj} = 3.0 \times 10^{14}$~cm, $\alpha_B^{\sj} = -1$, 
and $\beta_B^{\sj} = -2.5$.
The intrinsic spectral parameter $\gamma h{\nu'_0}^{\sj}$ and
the amplitude $A^{\sj}$ are determined so that
the time-averaged emission from a single subjet viewed {\it on-axis} satisfies
the following correlation,
\begin{equation}
\frac{L_{iso}^s}{10^{52}~\rm{erg}~\rm{s}^{-1}} =
\xi \left(\frac{E_p^s}{1~\rm{keV}}\right)^2,
\label{eq:yone}
\end{equation}
where $L_{iso}^s$ is the time-averaged ``bolometric''
isotropic-equivalent luminosity and
$E_p^s$ is the time-averaged rest-frame spectral peak energy
of the {\it on-axis} emission from a single subjet.
As for the validity of this correlation,
\citet{liang04} argue that for long bright BATSE GRBs
the observed $\gamma$-ray flux $F$ is correlated
with the observed time-resolved $E_p^{\obs}$ at each time in a similar way, i.e.,
$F \propto ({E_p^{\obs}})^2$, which supports the assumption that the on-axis emission
of each subjet obeys this correlation.
\citet{lloyd02} show that there is a positive correlation between 
$\gamma$-ray luminosity and time-resolved rest-frame spectral peak energy 
by using variability-luminosity correlation \citep[see also][]{yone04,ghir05b}.
This correlation could be obtained by standard synchrotron internal shock model 
\citep[e.g.,][]{zhang02b}.
However, the coefficient $\xi$ is highly uncertain.
Therefore we chose the values of $\xi$ so that the results of simulations 
reproduce the observations.
We consider two cases of $\gamma h{\nu'_0}^{\sj}$ and $A^{\sj}$:
Case (i) $\gamma h{\nu'_0}^{\sj}$ and $\xi$ are fixed as 350 keV and
$6.0 \times 10^{-5}$, respectively, for all $j$.
Case (ii) $\gamma h{\nu'_0}^{\sj}$ and $\xi$ are distributed around the
above values.

\subsection{Case (i)}

Let us consider the Case (i) as a simple toy model,
in which all the subjets have the same intrinsic parameters, so that
we can investigate the pure kinematical
effects from the multiple discrete emission patches.
The results are shown in Figure~\ref{fig:dap}.
The black solid line shows the $E_p$--$E_{iso}$ correlation
for a single subjet derived with
the same parameters. 
We see that the black solid line traces
the left-side edge of the
distribution of the simulated bursts.
When a single subjet is seen on-axis, 
the time-averaged spectral peak energy $E_p^s = g_n h \nu'_0 /
\gamma(1-\beta) \approx 2 g_n \gamma h \nu'_0 \sim 500$~keV,
where a numerical factor $g_n (\sim 0.7)$ comes from the contribution
of soft emission from the whole subjet, while $g_n =1$ in the case of
point source approximation.
The observed pulse has a duration determined by
the angular spreading time as
$\delta T = r_0 {\Delta\theta_{\sub}}^2/ 2c = 2$~s.
Then, according to equation (\ref{eq:yone}),
$L_{iso}^s \simeq 1.5 \times 10^{53}~\rm{ergs}~{s}^{-1}$, so that
$E_{iso}^s = L_{iso}^s \delta T \simeq 3 \times 10^{53}$~ergs.
This corresponds to $E_p$ reaching its maximum around 
$E_{iso} \sim 3 \times 10^{53}$~ergs.
When more than 1 subjet are seen on-axis, i.e., $n_s \geq 2$,
$E_p$ is the same as in the case of $n_s = 1$, but
$E_{iso} \simeq n_s E_{iso}^s$.
The maximum value of multiplicity $n_s$ is about 30, when
the line of sight is along the center of the whole jet. 
Then $E_{iso}$ takes the maximum value of $\simeq 10^{55}$~ergs.
Points with $E_p < 500$~keV correspond to the case 
of $n_s = 0$, in which all the subjets are seen off-axis, 
i.e., $\theta_v^{\sj} > \Delta\theta_{\sub}$ for all $j$.
For each line of sight, the observed flux is dominated by the emission of 
the subjets with small $\theta_v^{\sj}$.
Thus $E_p$ is determined by the minimum value of $\theta_v^{\sj}$, 
$\theta_v^{\rm min}$.
Let $n_s^{\rm off}$ be the number of the subjets with $\theta_v^{\sj}$ 
around $\theta_v^{\rm min}$.
When $n_s^{\rm off} = 1$, the observed flux is dominated by a single subjet,
and the $\theta_v^{\rm min}$-dependence of $E_p$ and $E_{iso}$
is determined as discussed in \S~\ref{sec:single}. Such points are
on the black solid line. 
When $n_s^{\rm off} \geq 2$, for each $\theta_v^{\rm min}$,
$E_p$ is the same as for the case of
$n_s^{\rm off} = 1$, but $E_{iso} \simeq n_s^{\rm off} E_{iso}^s.$
Thus the scatter of the simulated points for $E_p < 500$~keV
arises from that of $n_s^{\rm off}$.
We find that the right-side edge of the distribution of the points follows
$E_p \propto {E_{iso}}^{1/2}$. 
The reason for this behavior is as follows.
For each $\theta_v^{\rm min}$,  
$E_p \propto [1-\beta\cos(\theta_v^{\rm min} - 
\Delta\theta_{\sub})]^{-1}$.
The other quantity $E_{iso}$ is given for the largest $n_s^{\rm off}$.
Since the probability that these $n_s^{\rm off}$ subjets have the same axis 
$(\vartheta^{\sj}, \varphi^{\sj})$ is quite low, 
they should be smoothly distributed around the line of sight.
Then in calculating $E_{iso}$ by equation (\ref{eq:eiso})
for the multiple subjets case, 
we can take $\Delta\phi \simeq \pi$.
Therefore, for each $\theta_v^{\rm min}$, $E_{iso} \propto
[1-\beta\cos(\theta_v^{\rm min} - \Delta\theta_{\sub})]^{-2}$,
and then we obtain $E_{iso} \propto {E_p}^2$.
Such situation resembles the case of the annulus jet model in which
the line of sight is inside the annulus and the inner radius of the annulus 
changes \citep[see][]{eichler04}.

For a multiple subjet model, off-axis events (with $n_s =0$) follows
$E_p \propto {E_{iso}}^a$ with $0.4 < a < 0.5$.
This range of $a$ is obtained irrespective of the intrinsic subjet parameters
$\Delta \theta_{\sub}$, $\gamma$, $t_{\dep}$, $r_0$, 
$\alpha_B$, $\beta_B$, $\nu'_0$, and $A$.

\subsection{Case (ii)}

We here assume
that $\gamma h{\nu'_0}^{\sj}$ is distributed randomly
according to
a lognormal distribution function \citep{ioka02}
with an average of $\log (350~{\rm keV})$
and a logarithmic variance of 0.2.
For given $\gamma h{\nu'_0}^{\sj}$, $A^{\sj}$ is determined
by equation (\ref{eq:yone}).
The coefficient $\xi$ is also assumed to obey a lognormal distribution with
an average of $-5+ \log(6.0)$ and a logarithmic variance of 0.15.
The other parameters of the subjets are fixed to the same values as in the
previous simulation.
We calculate $E_{iso}$ and $E_p$, and then assign a redshift for
each
observer to calculate the distance and the observed lightcurve.
The source redshift distribution is assumed to be in proportion to the 
cosmic star formation rate.
We adopt the model SF2 in \citet{porci01}, in which we take the standard
cosmological
parameters of $H_0 = 70~\rm{km}~\rm{s}^{-1}~\rm{Mpc}^{-1}$, $\Omega_M =
0.3$,
and $\Omega_{\Lambda} = 0.7$.
Finally, we select detectable events with  observed peak photon
fluxes in the $1-10^4$~keV band larger than
$1.0~\rm{ph}~{\rm{cm}}^{-2}~{\rm{s}}^{-1}$,
which corresponds to the threshold sensitivity of {\it HETE-2}
\citep[see][]{band03,lamb05}.
Figure~\ref{fig:dap2} shows the result of our simulation.
Plus signs represent bursts that can be detected by {\it HETE-2},
while crosses represent ones that cannot be detected.
They are compared with the {\it BeppoSAX} and {\it HETE-2}
data (points with error bars) taken from \citet{ghir04}.
The solid line represents the best fitted line for 442 GRBs
with redshifts estimated by the lag--luminosity
correlation \citep{ghir05a}. We see that our simulated GRBs cover
the observed GRBs over three orders of $E_p$,
so that our multiple subjet model with the intrinsic correlation
$E_p^s \propto {L_{iso}^s}^{1/2}$
under the universal structured jet model 
is consistent with the observations.

The $n_s = 1$ bursts directly reflect the assumed correlation of 
$E_p^s \propto {L_{iso}^s}^{1/2}$.
For larger $n_s$, $E_{iso}$ becomes larger, and 
$E_p$ is determined
by the subjet emission with the largest $L_{iso}^s$ observed.
As a result, all the simulated bursts roughly obeys $E_p
\propto {E_{iso}}^{1/2}$
over about three orders of magnitude in $E_p$.
The scatter comes from the differences of
the number of the observed subjets and the differences of the
parameters of each subjet.

\section{Discussion}
\label{sec:dis}

We have investigated $E_p$--$E_{iso}$ correlation in
a multiple subjet model for GRBs, XRRs, and XRFs.
We find that off-axis events (with $n_s =0$) for multiple discrete 
emission regions show $E_p \propto {E_{iso}}^a$ with $0.4 < a < 0.5$.
It is assumed that the subjet parameters 
$\gamma h{\nu'_0}^{\sj}$ and $A^{\sj}$ are distributed so that 
emission of the subjets viewed on-axis follows the 
correlation $E_p^s \propto {L_{iso}^s}^{1/2}$,
with narrow $E_p^s$ range (one order of magnitude).
Then the Amati correlation ($E_p \propto {E_{iso}}^{1/2}$) is reproduced 
over three orders of magnitude in $E_p$.
Although the scatter around the Amati correlation is large in the simulation,
the results are consistent with the observed properties of 
GRBs with known redshifts and the BATSE GRBs with pseudo redshifts 
derived from the lag-luminosity correlation.
We argue that for brighter bursts the Amati correlation arises from
intrinsic property, while for dimmer bursts it arises from 
the off-axis effects of multiple emissions.
The intrinsic $E_p^s \propto {L_{iso}^s}^{1/2}$ correlation is supported by the 
observations \citep{liang04,lloyd02} 
and could be derived in the context of standard synchrotron internal shock model.

\begin{figure}
\plotone{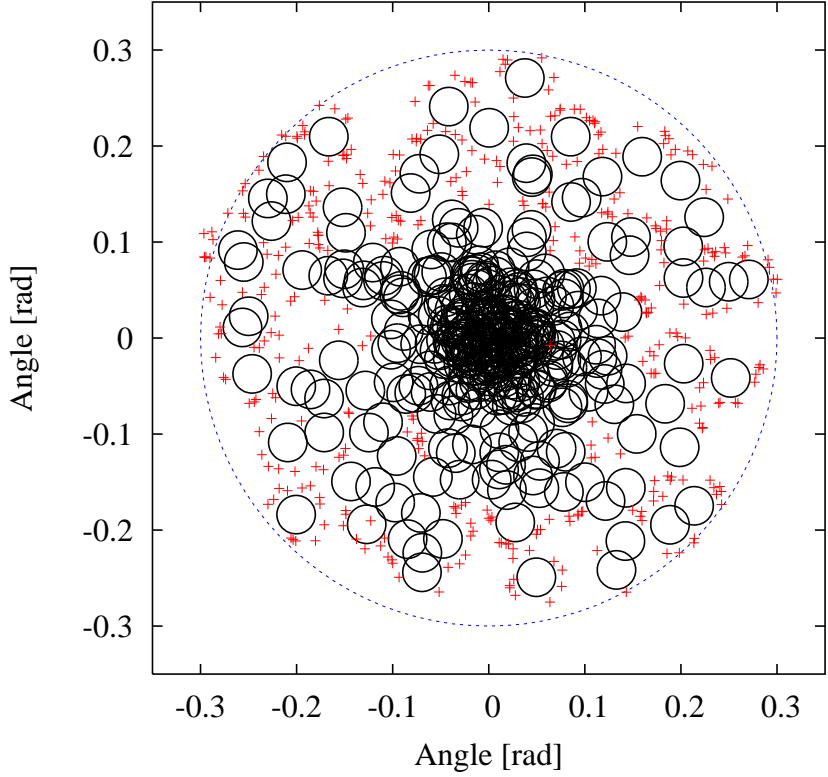}
\caption{
Angular distribution of $N_{\tot} = 350$ subjets confined in the whole
GRB jet in our simulation.
Each subjet is located according to the power-law distribution function
of eq.(\ref{eq:jetP}).
The whole jet has an opening half-angle of $\Delta\theta_{\tot} = 0.3$~rad.
The subjets have the same opening half-angles of
$\Delta\theta_{\sub} = 0.02$~rad.
The angular size of the subjets are represented by the solid
circles, while the whole jet is represented by the dashed circle.
The viewing angles for detectable XRFs in our simulation are
represented by plus signs.
}
\label{fig:jetP}
\end{figure}

\begin{figure}
\plotone{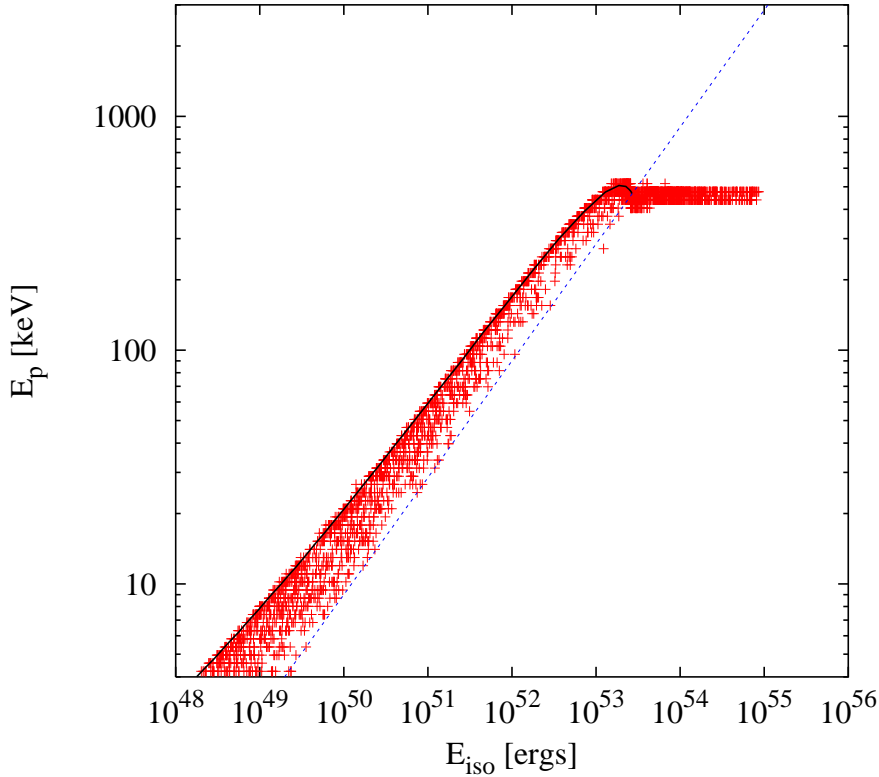}
\caption{
$E_p$--$E_{iso}$ diagram in the multiple subjet model in which all the
properties
of each subjet are the same.
The simulated bursts are represented by plus signs.
$E_p$--$E_{iso}$ line for a single subjet is described by solid line.
The dashed line is $(E_p/1~\rm{keV}) = 90(E_{iso}/10^{52}~\rm{ergs})^{1/2}$.
}
\label{fig:dap}
\end{figure}

\begin{figure}
\plotone{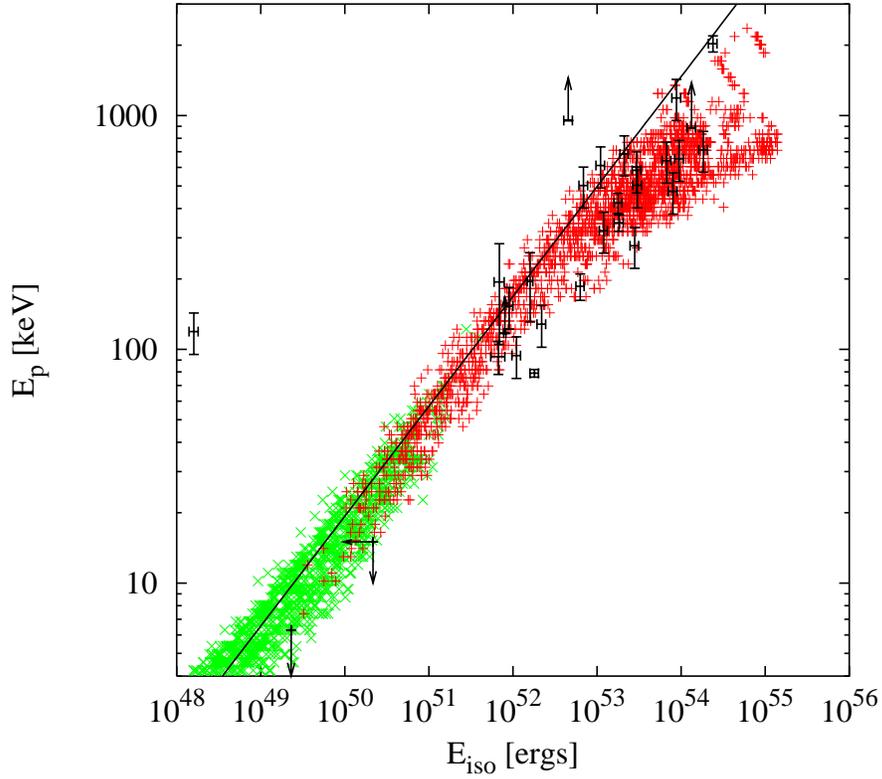}
\caption{
Same as Figure~\ref{fig:dap}, but in the multiple subjet model in which
subjet parameters $\gamma {\nu'_0}^{\sj}$, $A^{\sj}$, and $\xi$ are
distributed (see text for details).
Plus signs represent bursts that can be detected by {\it HETE-2},
while crosses represent ones that cannot be detected.
They are compared with the {\it BeppoSAX} and {\it HETE-2} data ({\it points
with error bars}) taken from \citet{ghir04}.
The solid line represents the best fitted line for 442 GRBs with
redshifts estimated by the lag--luminosity correlation \citep{ghir05a}.
}
\label{fig:dap2}
\end{figure}

\begin{figure}
\plotone{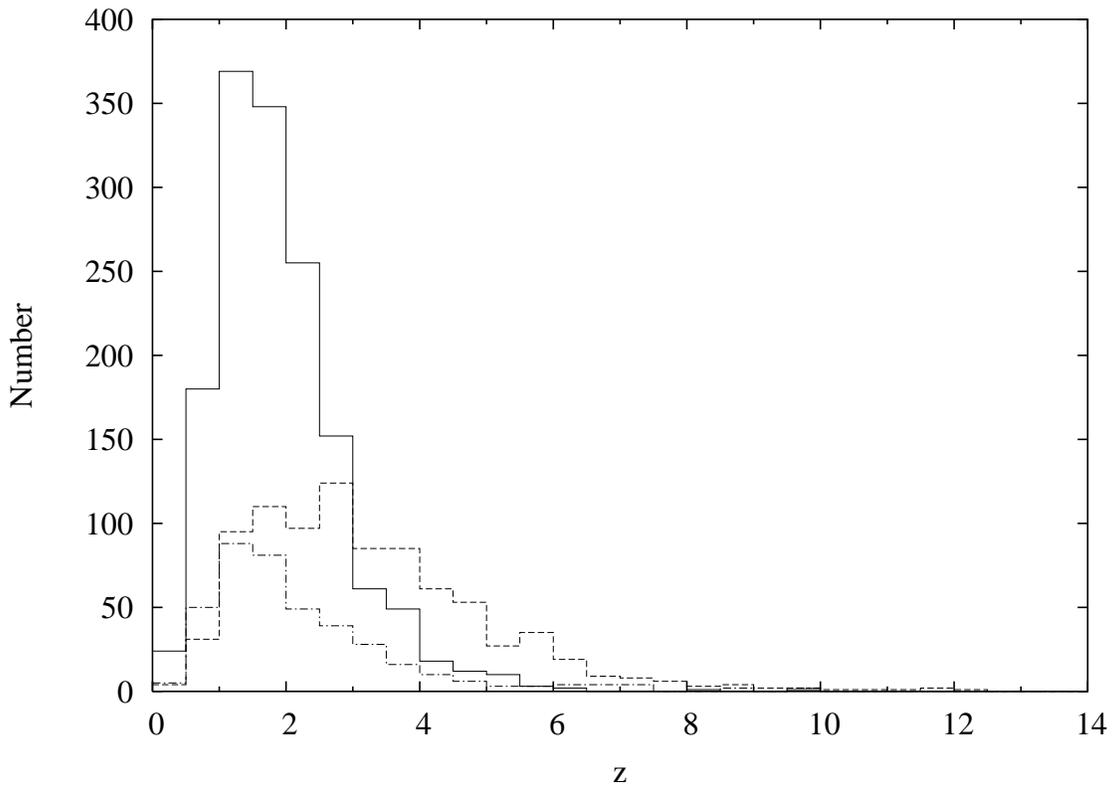}
\caption{
Redshift distribution of the simulated bursts surviving the peak
photon flux truncation.
Solid line, dashed line, and dot-dashed line represents the distribution for
GRBs, XRRs, and XRFs, respectively.
}
\label{fig:daz}
\end{figure}

{\it HETE} team defines XRRs and XRFs as those events for which
$\log[S_X (2-30~\rm{keV}) / S_{\gamma} (30-400~\rm{keV})] > -0.5$ and
$0.0$, respectively \citep{lamb04}.
We calculate the observed fluence ratio for simulated bursts surviving
the peak flux truncation, and classify them into GRBs, XRRs, and XRFs.
The ratio of the simulated event rate is
$R_{\rm{GRB}} : R_{\rm{XRR}} : R_{\rm{XRF}} \sim 4 : 3 : 1$.
{\it HETE-2} observations show similar number of GRBs, XRRs,
and XRFs \citep{saka05}. 
We can say that the event rate among GRBs, XRRs
and XRFs is consistent with the observations.
Figure~\ref{fig:daz} shows the redshift distribution
of GRBs (the solid line), XRRs (the dashed line)
and XRFs (the dot-dashed line). 
The mean redshifts of GRBs, XRRs and XRFs are 
1.9, 3.2, and 2.3, respectively.
XRRs have a little larger redshifts than GRBs and XRFs.
Figure~\ref{fig:jetP} plots the viewing angles for detectable XRFs,
which are represented by plus signs.
We see that
the main population of the XRFs arises from
the off-axis effects.
On the other hand, many XRRs are on-axis events.
Since $E_p > 200$~keV for on-axis events in our simulation,
the on-axis XRRs arise from the cosmological redshift effect.
The ratio of the on-axis and off-axis XRRs is $\sim$ 1 : 1.
We expect that the event rate ratio from larger
observed samples will give us some information
about the angular distribution of the subjets
within the whole GRB jet and the redshift distribution of the GRB sources.

In this paper,
we have performed the simulations with fixed Lorentz
factor of the subjets, $\gamma = 300$.
As discussed in \S~\ref{sec:multi}, the range of the index $a$ of 
the $E_p$--$E_{iso}$ correlation for off-axis events 
is independent of the Lorentz factor.
We perform the same simulations in the case (ii) for $\gamma=100$ and 
500, and obtain the Amati correlation ($E_p \propto {E_{iso}}^{1/2}$)
through all bursts.
However, the peak photon flux of the XRF
is small for lower Lorentz factor.
For $\gamma = 100$, we obtain
$R_{\rm{GRB}}: R_{\rm{XRR}} : R_{\rm{XRF}} \sim$ 15 : 10 : 1.
Alternatively for $\gamma=500$, 
$R_{\rm{GRB}}: R_{\rm{XRR}} : R_{\rm{XRF}} \sim$ 3 : 2 : 1.

\begin{figure}
\plotone{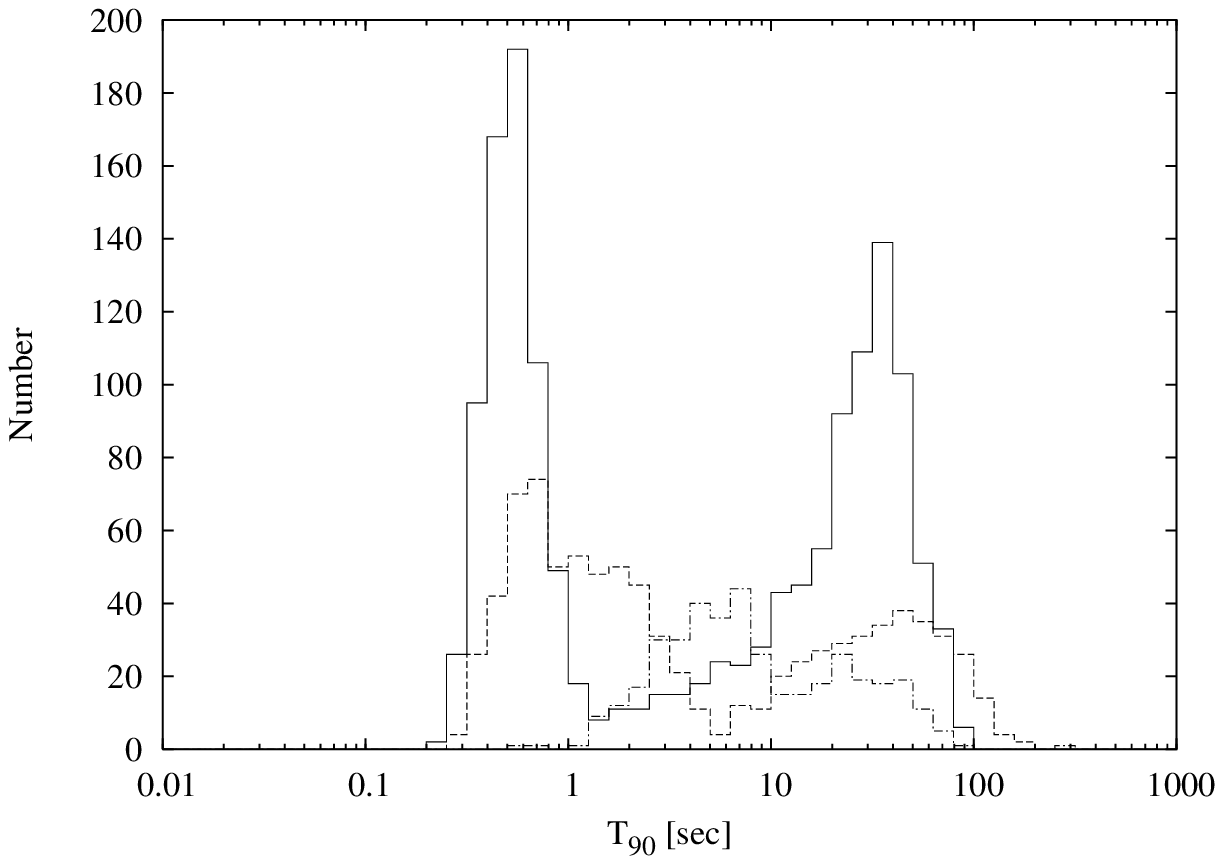}
\caption{
Distribution of the $T_{90}$ durations in the 50--300~keV band of the
simulated bursts surviving the peak flux truncation.
Solid line, dashed line, and dot-dashed line represents the distribution for
GRBs, XRRs, and XRFs, respectively.
}
\label{fig:da1}
\end{figure}

\begin{figure}
\plotone{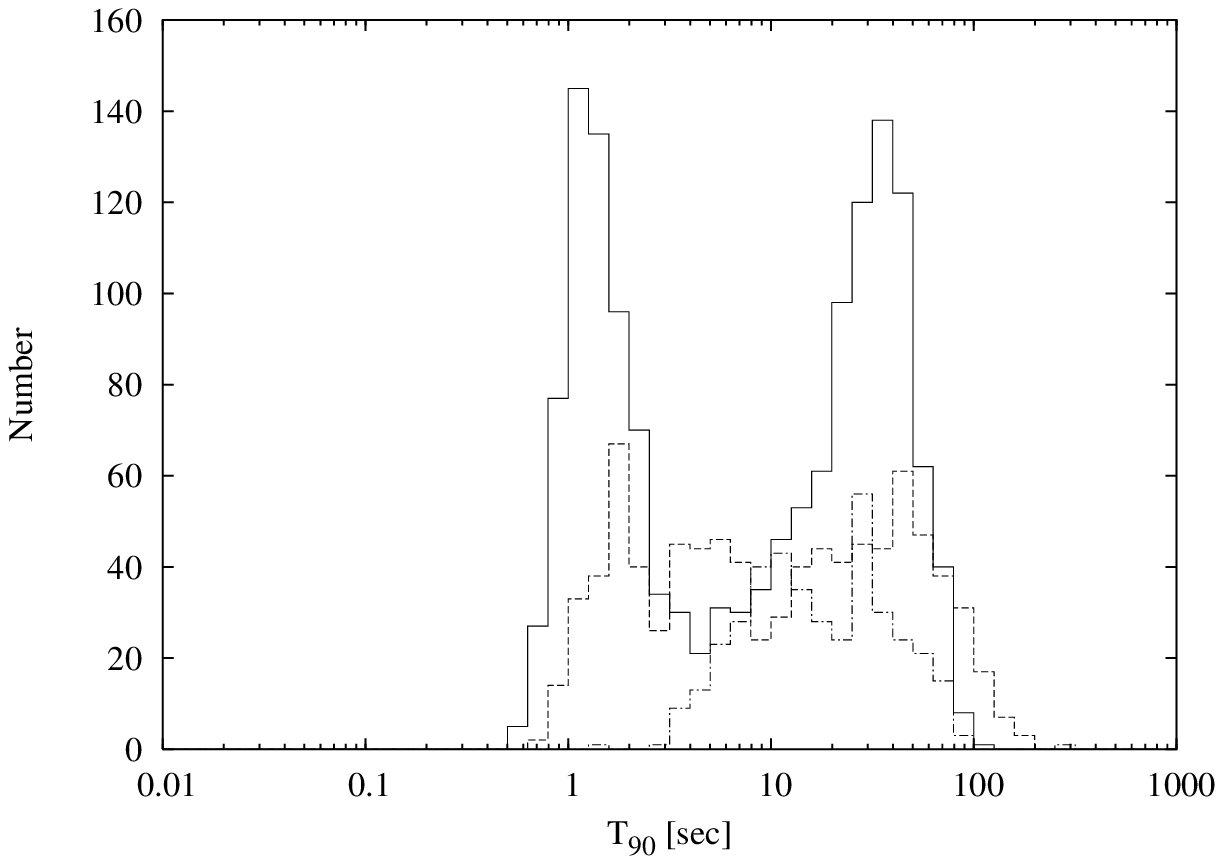}
\caption{
Distribution of the $T_{90}$ durations in the 2--25~keV band of the
simulated bursts surviving the peak flux truncation.
Solid line, dashed line, and dot-dashed line represents the distribution for
GRBs, XRRs, and XRFs, respectively.
}
\label{fig:da2}
\end{figure}

Figure~\ref{fig:da1} shows the distribution of
the $T_{90}$ durations
in the 50--300~keV band for GRBs (solid line), XRRs (dashed line),
and XRFs (dot-dashed line).
GRBs have a bimodal distribution as observed by BATSE.
We have already shown why GRBs have the bimodal
duration distribution
in our multiple subjet model \citep{toma05}:
the $T_{90}$ duration of an $n_s = 1$ burst is determined
by the width of a single pulse, while that of
an $n_s \geq 2$ burst is determined by the time interval
between the observed first pulse and the last one.
These two different timescales naturally lead
to a division of the burst
$T_{90}$ durations into the short and long ones.
We also calculate the distribution of the $T_{90}$ durations in the
2--25~keV band.
Figure~\ref{fig:da2} shows the result. 
These distributions are not inconsistent with the {\it HETE-2} data
\citep[see Fig.~4 of][]{saka05}.
The $T_{90}$ durations of $n_s =1$ bursts (i.e., short bursts)
become larger when they are measured in the softer band,
since soft emission from the periphery of the subjet is observed for
a longer time.
\citet{yama04b} have predicted short XRRs in our unified model,
which are confirmed in this simulation.
These are events of a single subjet viewed off-axis or viewed on-axis
with slightly high redshift.
Indeed, GRB 040924 may be an example of short XRRs, from which recent 
HST observation reveals the supernova signature \citep{soder05}.
This event supports our unified picture.

In this paper, we considered the $\theta^{-2}$-angular distribution
of the subjets.
Averaging by a solid angle satisfying
$(\Delta\theta_{\sub})^2<\Omega<(\Delta\theta_{\tot})^2$, the
distribution of the emission energy (or almost equivalently
the angle-averaged kinetic energy) is the same
as the universal structured jet model
\citep{rossi02,zhang02a}.
The universal structured jet model has been criticized
by \citet{lamb05}:
in the universal structured jet model, it is assumed that
XRFs are observed when the jet is viewed from fairly
large angle, so that the model overpredicts
the number of XRFs, which is inconsistent with the observed
ratio of the number of XRFs and GRBs detected by {\it HETE-2}.
Then, \citet{zhang04} modified the universal structured jet model,
and showed that if the jet is structured with
a Gaussian-like shape, the number of XRFs becomes small.
In these works it is assumed that the jet is continuous
and there are no cold spots inside the jet.
As shown in this paper, \citet{eichler04}, and \citet{yama04b},
if the observer points toward the cold spot (i.e., $n_s=0$),
XRFs or XRRs are observed. While if $n_s \geq 2$, the event looks
like a long GRB irrespective of the viewing angle. In our
model, the ratio of the total solid angle
with $n_s \geq 2$ and $n_s=0$
determines the event rate of GRBs and XRRs/XRFs.
Interestingly, we find that the power-law profile with an index
of $-2$ is preferable to the Gaussian profile
in order to reproduce the ratio of observed event
rate of GRBs, XRRs, and XRFs, because the
solid angle with $n_s=0$ is small in the Gaussian profile.
\citet{lazzati05} have recently argued
that in the context of the collapsar model,
$\theta^{-2}$ angular profile might be obtained
as a consequence of the physics in the jet breakout
irrespective of the jet structure inside the progenitor.
From the observational side, we can estimate the pseudo
jet opening angle distribution.
Using the Ghirlanda correlation ($E_p \propto {E_{\gamma}}^{0.71}$)
where $E_{\gamma} = E_{iso} {\theta_j}^2/2$ \citep{ghir04} and
the Yonetoku correlation ($E_p \propto {L_p}^{0.5}$) where
$L_p$ is a peak luminosity \citep{yone04},
\citet{yone05} obtained that the
pseudo jet opening angle obeys $f( \theta_j )d \theta_j
\propto \theta_j^{-2}d \theta_j$.
This is compatible with the power-law structured jet model:
if all bursts were
observable, the distribution would be uniform
per unit solid angle and $f(\theta) \propto \theta$. However
$E_{iso}$ for the
smaller viewing angle is brighter by a factor of $\theta^{-2}$,
so that the maximum observable distance is larger by a
factor of $\theta^{-1}$ which contains a volume larger by a factor of
$\theta^{-3}$. Then we have $f(\theta)\propto\theta^{-2}$.

Late phase evolution of a set of multiple subjets is rather complicated
and hard to be predicted.
Cold spots do not produce high energy emission but may be filled with
the kinetic energy that is not dissipated at small radius
\citep[see also][]{levinson05}.
Even if cold spots are not filled with the kinetic energy,
all subjets begin to expand sideways and would merge
into one shell.
In any case, late afterglow behavior may be well approximated
by the results from the continuous structured jet model
\citep[e.g.,][]{kumar04}.
As shown in Fig.~\ref{fig:jetP}, almost all XRFs arise
when all the subjets are viewed off-axis, i.e., $n_s=0$,
while the observers see the whole jet on-axis.
Then, the late phase ($\gtrsim1$~day) properties of
XRF afterglows may
not be like orphan afterglows but may show
similar behavior to those of normal GRBs \citep[e.g.,][]{amati04}.
On the other hand, as rare cases,
when the whole jet is viewed off-axis,
XRF afterglows may resemble the orphan afterglow
\citep[e.g.,][]{granot05}.
XRF~030723 may be a member of such a class
\citep{butler05,fynbo04}.

\acknowledgements
We are grateful to the referee for useful comments.
K.~T. thanks M.~Ohashi for helpful discussion.
This work is supported in part by
the Grant-in-Aid for the 21st Century COE
``Center for Diversity and Universality in Physics''
from the Ministry of Education, Culture, Sports, Science and Technology
(MEXT) of Japan
and also by Grants-in-Aid for Scientific Research
of the Japanese Ministry of Education, Culture, Sports, Science,
and Technology 09245 (R.~Y.), 14047212 (T.~N.), 14204024 (T.~N.)
and 17340075 (T.~N.).

\clearpage

\begin{thebibliography}{99}

\bibitem[Amati et al.(2002)]{amati02}
Amati,~L. et al. 2002, A\&A, 390, 81
%
\bibitem[Amati et al.(2004)]{amati04}
Amati,~L. et al. 2004, A\&A, 426, 415
%
\bibitem[Band et al.(1993)]{band93}
Band,~D.~L., et al. 1993, ApJ, 413, 281
%
\bibitem[Band et al.(2003)]{band03}
Band,~D.~L. 2003, ApJ, 588, 945
%
\bibitem[Butler et al.(2005)]{butler05}
Butler,~N.~R., et al. 2005, ApJ, 621, 884
%
\bibitem[Eichler \& Levinson(2004)]{eichler04}
Eichler,~D., \& Levinson,~A. 2004, ApJ, 614, L13
%
\bibitem[Fynbo et al.(2004)]{fynbo04}
Fynbo,~J.~P.~U., et al. 2004, ApJ, 609, 962
%
\bibitem[Ghirlanda, Ghisellini, \& Lazzati(2004)]{ghir04}
Ghirlanda,~G., Ghisellini,~G., \& Lazzati,~D. 2004, ApJ, 616, 331
%
\bibitem[Ghirlanda, Ghisellini, \& Firmani(2005a)]{ghir05a}
Ghirlanda,~G., Ghisellini,~G., \& Firmani,~C. 2005a, MNRAS, 361, L10
%
\bibitem[Ghirlanda et al.(2005b)]{ghir05b}
Ghirlanda,~G., Ghisellini,~G., Firmani,~C., Celloti,~A., \&
Bosnjak,~Z. 2005b, MNRAS, 360, L45
%
\bibitem[Granot, Ramirez-Ruiz, \& Perna(2005)]{granot05}
Granot,~J., Ramirez-Ruiz, E., \& Perna, R., 2005, preprint
(astro-ph/0502300)
%
\bibitem[Ioka \& Nakamura(2001)]{ioka01}
Ioka,~K., \& Nakamura~T. 2001, ApJ, 554, L163
%
\bibitem[Ioka \& Nakamura(2002)]{ioka02}
Ioka,~K., \& Nakamura,~T. 2002, ApJ, 570, L21
%
\bibitem[Kumar \& Piran(2000)]{kumar00}
Kumar,~P., \& Piran,~T. 2000, ApJ, 535, 152
%
\bibitem[Kumar \& Granot(2004)]{kumar04}
Kumar,~P. \& Granot,~J. 2004, ApJ, 591, 1075
%
\bibitem[Lamb et al.(2004)]{lamb04}
Lamb,~D.~Q., et al. 2004, NewA Rev., 48, 423
%
\bibitem[Lamb, Donaghy, \& Graziani(2005)]{lamb05}
Lamb,~D.~Q., Donaghy,~T.~Q., \& Graziani,~C. 2005, ApJ, 620, 355
%
\bibitem[Lazzati \& Begelman(2005)]{lazzati05}
Lazzati,~D. \& Begelman,~M.~C. 2005, preprint (astro-ph/0502084)
%
\bibitem[Levinson \& Eichler(2005)]{levinson05}
Levinson,~A. \& Eichler,~D. 2005, preprint (astro-ph/0504125)
%
\bibitem[Liang, Dai, \& Wu(2004)]{liang04}
Liang,~E.~W., Dai~Z.~G., \& Wu,~X.~F. 2004, ApJ, 606, L29
%
\bibitem[Lloyd-Ronning \& Ramirez-Ruiz(2002)]{lloyd02}
Lloyd-Ronning,~N.~M., \& Ramirez-Ruiz,~E. 2002, ApJ, 576, 101
%
\bibitem[Nakamura(2000)]{nakamura00}
Nakamura,~T. 2000, ApJ, 534, L159
%
\bibitem[Porciani \& Madau(2001)]{porci01}
Porciani,~C., \& Madau,~P. 2001, ApJ, 548, 522
%
\bibitem[Rossi, Lazzati, \& Rees(2002)]{rossi02}
Rossi,~E., Lazzati,~D., \& Rees,~M.~J. 2002, MNRAS, 332, 945
%
\bibitem[Sakamoto et al.(2005)]{saka05}
Sakamoto,~T., et al. 2005, ApJ, 629, 311
%
\bibitem[Soderberg et al.(2005)]{soder05}
Soderberg,~A.~M., et al. 2005, preprint (astro-ph/0504359)
%
\bibitem[Toma, Yamazaki, \& Nakamura(2005)]{toma05}
Toma,~K., Yamazaki,~R., \& Nakamura,~T. 2005, ApJ, 620, 835
%
\bibitem[Woods \& Loeb(1999)]{woods99}
Woods,~E. \& Loeb,~A. 1999, ApJ, 523, 187
%
\bibitem[Yamazaki, Ioka, \& Nakamura(2002)]{yama02}
Yamazaki,~R., Ioka,~K., \& Nakamura,~T. 2002, ApJ, 571, L31
%
\bibitem[Yamazaki, Ioka, \& Nakamura(2003)]{yama03}
Yamazaki,~R., Ioka,~K., \& Nakamura,~T. 2003, ApJ, 593, 941
%
\bibitem[Yamazaki, Ioka, \& Nakamura(2004a)]{yama04a}
Yamazaki,~R., Ioka,~K., \& Nakamura,~T. 2004a, ApJ, 606, L33
%
\bibitem[Yamazaki, Ioka, \& Nakamura(2004b)]{yama04b}
Yamazaki,~R., Ioka,~K., \& Nakamura,~T. 2004b, ApJ, 607, L103
%
\bibitem[Yonetoku et al.(2004)]{yone04}
Yonetoku,~D., Murakami,~T., Nakamura,~T., Yamazaki,~R., Inoue,~A.~K., \&
Ioka,~K. 2004, ApJ, 609, 935
%
\bibitem[Yonetoku et al.(2005)]{yone05}
Yonetoku,~D., Yamazaki,~R., Nakamura,~T., \& Murakami,~T. 2005, 
preprint (astro-ph/0503254)
%
\bibitem[Zhang \& M\'{e}sz\'{a}ros(2002a)]{zhang02a}
Zhang,~B., \& M\'{e}sz\'{a}ros,~P. 2002a, ApJ, 571, 876
%
\bibitem[Zhang \& M\'{e}sz\'{a}ros(2002b)]{zhang02b}
Zhang,~B., \& M\'{e}sz\'{a}ros,~P. 2002b, ApJ, 581, 1236
%
\bibitem[Zhang et al.(2004)]{zhang04}
Zhang,~B., Dai,~X., Lloyd-Ronning,~N.~M., \& M\'{e}sz\'{a}ros,~P. 2004, ApJ, 601, L119
%
\end{thebibliography}
\end{document}